\documentclass[preprint,showpacs]{revtex4}
\usepackage{epsfig}
\usepackage{graphicx}
\usepackage{slashed}
\usepackage{bm}
\usepackage{amssymb}
\newcommand{\half}{\mbox{${\textstyle \frac{1}{2}}$}}           % 1/2
\newcommand{\fourth}{\mbox{${\textstyle \frac{1}{4}}$}}         % 1/4
\newcommand{\rd}{\textrm{d}}
\begin{document}
%{\flushright{\today}}
\title{Entanglement in joint $\Lambda \bar{\Lambda}$ decay}

\date{\today}
\author{G\"oran F\"aldt}\email{goran.faldt@physics.uu.se}  
\affiliation{ Department of physics and astronomy, 
Uppsala University,
 Box 516, S-751 20 Uppsala,Sweden }

\begin{abstract}
We investigate the joint $\Lambda \bar{\Lambda}$ decay in the reaction 
$e^+ e^- \rightarrow \gamma \Lambda(\rightarrow p\pi^-) \bar{\Lambda}(\rightarrow \bar{p}\pi^+)$.
This reaction 
%%is interesting since it 
may
provide information on the electromagnetic form factors of the
Lambda baryon, in the time-like region.  We present a conventional diagram-based calculation where
production and decay steps are coherent and summations over final-state proton
and anti-proton spins are performed. The resulting cross-section distribution is explictly
covariant as it is expressed in  scalar products of the four-momentum vectors of
the participating particles. 
We compare this calculation with that of the folding method which we extend and make
explicitly covariant. In the folding method production and decay distributions, 
not amplitudes, are folded together. Of particular importance is then  a correct couting of the 
number of possible intermediate-hyperon-spin states.

\end{abstract}
\maketitle
%
%
%%%%%%%%%%%%%%%%%%%%%%%%%%%
%
\section{Introduction}\label{ett}

The BaBar detector \cite{BaBar} has been used to study a number of $e^+e^-$ annihilation 
reactions.  One of them is the initial-state-radiation reaction,   
$e^+ e^- \rightarrow \gamma \Lambda \bar{\Lambda}$, which  offers
means to determine
the electromagnetic form factors of the $\Lambda$ hyperon in
the time-like region. This determination is achieved by varying
the energy of the radiated photon. 

A theoretical analysis of the above reaction is presented in  Refs.\cite{Novo} and \cite{Czyz}. 
It is based on the folding method. The cross-section distribution is obtained 
by multiplying distributions functions for the $\Lambda\bar{\Lambda}$ production with the decay-distribution
functions for the Lambda and anti-Lambda hyperons, all for  fixed hyperon-spin 
directions. This product is then averaged over the hyperon-spin directions. The disadvantage
of this method, as used, is that several coordinate systems are employed in the 
calculation, and there seems to be a problem of properly counting the number of intermediate
hyperon-spin states. The method raises some doubts since the product distributions with 
fixed intermediate-hyperon-spin directions are unphysical. 

We prefer the conventional perturbation method, calculating directly the relevant 
diagram-matrix elements
that automatically sum over the spin components of the intermediate hyperons. 
This yields  cross-section distributions which are explicitly covariant as
they are expressed in terms of scalar products of the four momenta of the 
participating particles, which means directly in terms of the measured four momenta.

We also perform  the calculations using the folding method, which we first extend
into a covariant method. A comparison with the conventional diagrammatic method
shows that they give identical results, provided the possible intermediate-hyperon-spin states are properly counted.
Both methods involve about the same calculational effort.

\newpage
%
%%%%%%%%%%%%%%%%%%%%%%%%%%%
%
\section{Lambda form factors}\label{två}

The two diagrams under consideration are graphed in Fig.1. Our momentum definitions
are also indicated there. In the diagrams the decays of the $\Lambda$ hyperons are
included, with decay vertices as defined in  Appendix A. The coupling of the
initial state leptons are simply given by the electron charge. No form factors or 
anomalous magnetic moments for the leptons are considered.
\begin{figure}[ht]
\scalebox{0.65}{\includegraphics{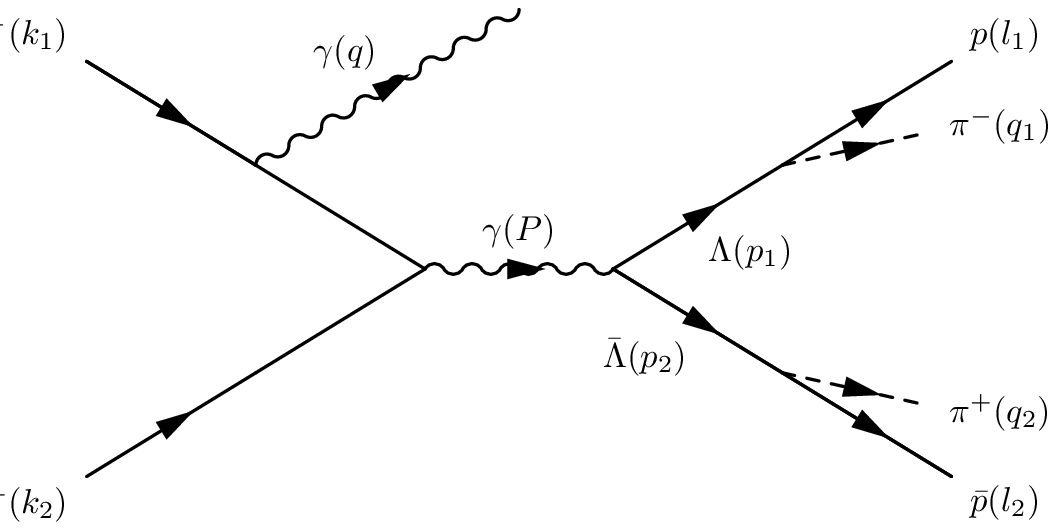}\qquad \includegraphics{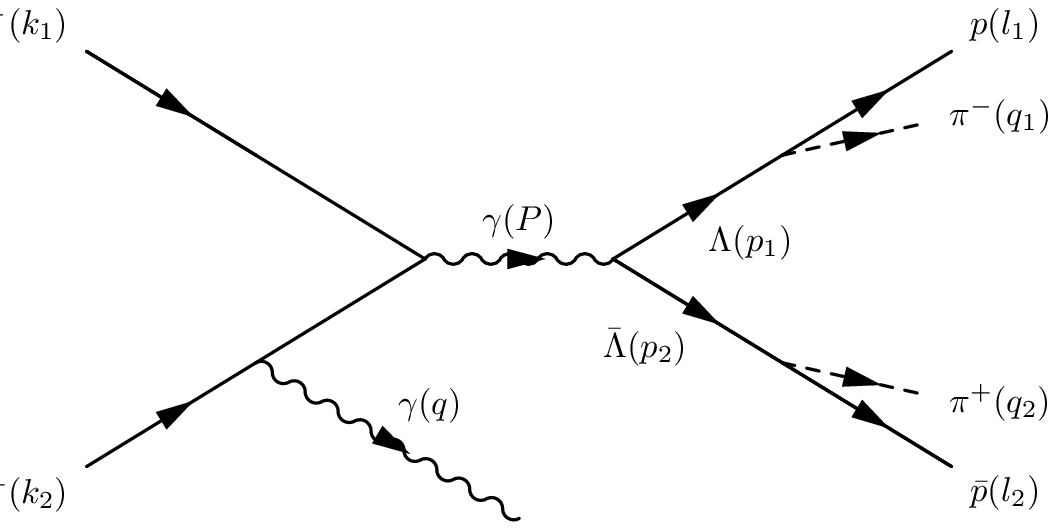}  }
\caption{Graphs included in our calculation of the reaction 
$e^+ e^- \rightarrow \gamma \Lambda(\rightarrow p\pi^-) \bar{\Lambda}(\rightarrow \bar{p}\pi^+)$.}
\label{F1-fig}
\end{figure}

In the current matrix elements of the $\Lambda$ hyperon, however, both form
factors and their momentum dependencies are taken into account. After all
this is what experiments aim to determine. It is common to write 
the hadron current matrix element as
\begin{eqnarray}
		j_\mu(p_1,p_2)&=&-ie\bar{u}(p_1)O_\mu(p_1,p_2)v(p_2) \\
		O_\mu(p_1,p_2)&=&G_1(P^2)\gamma_\mu-
		  \frac{1}{2M}G_2(P^2)Q_\mu \label{Lambdavertex}
\end{eqnarray}
with $P^2=(p_1+p_2)^2$ and $Q=p_1-p_2$.

The form factors $G_1$ and $G_2$ are related to the more commonly used form factors $F_1$ and $F_2$,
and the electric $G_E$ and magnetic 
$G_M$ form factors \cite{Czyz,Dalk,Pil}, through
\begin{eqnarray}
	G_1&=& F_1+F_2 = G_M	  \\
	G_2&=&F_2=\frac{1}{1+\tau}(G_M-G_E) =\frac{4M^2}{Q^2}(G_M-G_E), 
\end{eqnarray}
and $\tau=-P^2/4M^2$. The arguments of the form factors are all equal to $P^2$. In particular, when 
$P^2=4M^2$ then $G_M=G_E=F_1+F_2$.

\newpage
%
%%%%%%%%%%%%%%%%%%%%%%%%%%%
%
\section{Cross section}\label{tre}

Our notation follows Pilkuhn \cite{Pil}. The cross-section distribution for the reaction
$e^+ e^- \rightarrow \gamma \Lambda(\rightarrow p\pi^-) \bar{\Lambda}(\rightarrow \bar{p}\pi^+)$
 is written as 
\begin{equation}
	\rd \sigma= \frac{1}{2\sqrt{\lambda(s,m_e^2,m_e^2)}} \, \overline{|{\cal{M}}|^2}\,
	   \textrm{dLips}(k_1+k_2;q,l_1,l_2,q_1,q_2)	  ,
\end{equation}	   
where the average over the squared matrix element indicates summation over final proton
and anti-proton spins and average over initial electron and positron spins.  The definitions of the
particle momenta are explained in Fig.1.

We would like to remove some trivial factors from the squared matrix element,
namely the powers of the electron charge and the squares of the intermediate  Lambda and anti-Lambda
denominators as well as the intermediate-photon denominator. These factors together give 
\begin{equation}
{\cal{K}}=	\frac{e^6}{(P^2)^2} \frac{1}{[(s_1-M^2)^2 +M^2\Gamma^2(\sqrt{s_1})][(s_2-M^2)^2 +M^2\Gamma^2(\sqrt{s_2})]},
  \label{Kdef}
\end{equation}
with $s_1=p_1^2$ and $s_2=p_2^2$. 
Since the hyperon widths are narrow,  $M\Gamma\ll 1$, they may be evaluated  at 
$\sqrt{s}=M$. Furthermore, it should be remembered that the width $\Gamma(M)=\Gamma(M;\Lambda\rightarrow all)$
 is the total Lambda (anti-Lambda) decay width.
 As a consequence of this factorization we may write
\begin{equation}
	\overline{|{\cal{M}}|^2}={\cal{K}}\overline{|{\cal{M}}_{red}|^2},\label{Msq_andK}
\end{equation}
with 
\begin{equation}
	{\cal{K}}= \frac{(4\pi\alpha)^3}{(P^2)^2}
	\frac{\pi^2}{M^2\Gamma^2(M)}\delta(s_1-M^2)\delta(s_2-M^2). \label{K-factor}
\end{equation}

Since the intermediate-hyperon states are states whose masses in the narrow-width approximation 
may be considered fixed,
it is useful to rewrite the phase-space expression making this explicit by using the
following nesting formula
\begin{eqnarray}
	 \textrm{dLips}(k_1+k_2;q,l_1,l_2,q_1,q_2)&=&\frac{1}{(2\pi)^2} \rd s_1 \rd s_2
		\textrm{Lips}(k_1+k_2;q,p_1,p_2)  \nonumber \\
		&& \times\textrm{dLips}(p_1;l_1,q_1)\textrm{dLips}(p_2;l_2,q_2),
		 \label{Ph_space_form}
\end{eqnarray}
with $p_1^2=s_1$  and $p_2^2=s_2$.
Multiplication by ${\cal{K}}$ puts the hyperons on their mass shells.
%
%%%%%%%%%%%%%%%%%%%%%%%%%%%
\newpage
\section{Lepton tensor}\label{fyra}
The leptonic four-current is defined as
\begin{equation}
	L_\mu(k_1,k_2,q)=\bar{v}(k_2)\gamma_\mu 
	 \frac{(\slashed k_1-\slashed q)+m_e}{(k_1-q)^2-m_e^2}\slashed \varepsilon u(k_1)
	 +\bar{v}(k_2)\slashed \varepsilon 
	 \frac{(-\slashed k_2+\slashed q)+m_e}{(k_1-q)^2-m_e^2}\gamma_\mu u(k_1),
	 \label{Lepton_vector}
\end{equation}
where index $\mu$ is tied to the lepton-intermediate-photon vertex.
For the cross-section distribution we need the corresponding leptonic tensor, 
\begin{equation}
	L_{\nu\mu}(k_1,k_2,q) = \fourth \sum L_\nu^{\dagger}(k_1,k_2,q) L_\mu(k_1,k_2,q),
\end{equation}
where the sum runs over initial lepton spins and  final 
photon polarizations. We neglect the electron mass  $m_e$ compared with other masses 
and energies. Furthermore, the lepton tensor enters  the cross-section distribution contracted
with the 
hadron tensor. The hadron tensor is gauge invariant, which means that when contracted
with four vectors $P^\mu$ or $P^\nu$ zero result is obtained. Hence, dependencies 
$P_\mu$ or $P_\nu$  in the lepton
tensor may be ignored. As a consequence, the relevant part of the lepton tensor becomes
symmetric in its indices and equal to
\begin{eqnarray}
	L_{\nu\mu}& = & L_{\mu\nu} \nonumber   \\
	 &=& \frac{1}{y_1y_2}\bigg[ -4(s-y_1-y_2)\{k_{1\nu}k_{1\mu}+ k_{2\nu}k_{2\mu}\}
	      \nonumber\\
	  && \qquad \qquad -\{2s(s-y_1-y_2) +y_1^2+y_2^2\}g_{\nu\mu} \bigg],\label{Lepton_tensor}
\end{eqnarray}
with
\begin{eqnarray}
	s & = & (k_1+k_2)^2,  \\
	 y_1 &=& -(k_1-q)^2 +m_e^2=2k_1    \cdot q,\\
	  y_2 &=& -(k_2-q)^2 +m_e^2 = 2 k_2\cdot q.          
\end{eqnarray}
We remark that
\begin{equation}
	s-y_1-y_2=(k_1+k_2-q)^2=P^2,
\end{equation}
and $P=p_1+p_2$. Our expression for the lepton tensor, Eq.(\ref{Lepton_tensor}), agrees 
with that of Czy\.z et al.\  \cite{Czyz}.

%%%%%%%%%%%%%%%%%%%%%%%%%%%
\newpage
\section{Hadron tensor}\label{fem}

The hadronic four-current $H_\mu(p_1,p_2,l_1,l_2)$ describes, in addition to the coupling of the
intermediate  photon
to the hyperons,  also their decays. The two parts are coherent. 
 The denominators of the hyperon propagators have already been extracted into the $\cal{K}$ factor
 of Eq.(\ref{Kdef}) so we are left with
\begin{eqnarray}
	H_\mu&=&  \bar{u}(l_1)[A+B\gamma_5] (\slashed p_1+M)  O_\mu(p_1,p_2) (\slashed p_2-M) 
    [A'+B'\gamma_5] v(l_2)
	 ,\label{Hadron_vector}
\end{eqnarray}
where the Lambda vertex function $O_\mu(p_1,p_2)$ is defined in Eq.(\ref{Lambdavertex}),
\begin{equation}
		O_\mu(p_1,p_2)=G_1(P^2)\gamma_\mu-
		  \frac{1}{2M}G_2(P^2)Q_\mu . \label{Omudef}
\end{equation}
The definition of the hadronic tensor is 
\begin{equation}
	H_{\nu\mu}=\sum H_\nu^\dagger H_\mu ,
\end{equation}
with the sum running over  final state proton and anti-proton polarizations.

The calculation of the hadronic tensor is simplified by noting that
\begin{equation}
	H_{\nu\mu}=\mbox{Sp}[Y_\Lambda \bar{O}_\nu X_\Lambda O_\mu], \label{hadron-tensor-trace}
\end{equation}
with $\bar{O}=\gamma_0 O^\dagger \gamma_0$, and
\begin{eqnarray}
	X_\Lambda&=&(\slashed p_1+M)\bigg[R_\Lambda-S_\Lambda \gamma_5(l_1\cdot p_1+M\slashed l_1)\bigg] ,
	  \label{LamXfactor}\\
	 Y_\Lambda&=&(\slashed p_2-M)\bigg[\bar{R}_\Lambda+
	          \bar{S}_\Lambda \gamma_5(l_2\cdot p_2-M\slashed l_2)\bigg] \label{anLamYfactor} .     
\end{eqnarray}
The $R$ and $S$ parameters govern the Lambda-hyperon decays and are defined in
Appendix A. We also note that the scalar products $l_1\cdot p_1=l_2\cdot p_2$ 
are constants.

We decompose the hadron tensor into powers of $R$ and $S$, writing
\begin{equation}
	H_{\nu\mu}=\bar{R}_\Lambda R_\Lambda H_{\nu\mu}^{RR}+\bar{R}_\Lambda S_\Lambda H_{\nu\mu}^{RS}
	  + \bar{S}_\Lambda R_\Lambda H_{\nu\mu}^{SR} +\bar{S}_\Lambda S_\Lambda H_{\nu\mu}^{SS}.\label{hadron-exp}
\end{equation}
The explicit expression for the first-partial-hadron tensor is the following,
\begin{eqnarray}
	H_{\nu\mu}^{RR}&=&2\bigg( (P_\nu P_\mu -P^2g_{\nu\mu})- Q_\nu Q_\mu  \bigg)|G_1|^2 \nonumber \\
	   &&+2Q_\nu Q_\mu\bigg(2\Re (G_1 G_2^{\star})
	  -\frac{Q^2}{4M^2}|G_2|^2\bigg).   \label{hOO}  
\end{eqnarray}
The argument of the form factors, which is $P^2$, is here omitted, and $P^2=4M^2-Q^2$. We also remark that
the two contributing terms are separately gauge invariant, i.e., they vanish upon contraction
with $P^\mu$ or $P^\nu$.

The terms involving spin contributions look like
\begin{eqnarray}
	H_{\nu\mu}^{RS} &=& - 4i |G_1|^2\bigg[l_1\cdot p_1\epsilon(p_2p_1)_{\nu\mu}
  - M^2\epsilon(p_1+p_2,l_1)_{\nu\mu} \bigg]\nonumber \\
	&&+2iG_1G_2^\star Q_\nu\epsilon(p_2p_1l_1)_\mu 
	   -2iG_1^\star G_2 Q_\mu\epsilon(p_2p_1l_1)_\nu ,   
\end{eqnarray}
with 
\begin{eqnarray}
	\epsilon(p_2p_1l_1)_\nu&=&\epsilon_{\alpha\beta\gamma\nu}p_2^\alpha p_1^\beta l_1^\gamma,\label{epsI} \\
	\epsilon(p_2p_1)_{\nu\mu}&=&\epsilon_{\alpha\beta\nu\mu}p_2^\alpha p_1^\beta ,
\end{eqnarray}
and $\epsilon_{0123}=1$. Now, we observe that the imaginary part of the tensor $H_{\nu\mu}^{RS}$ is anti-symmetric 
in its indices, whereas the real part is symmetric. Since the hadron tensor is to be contracted 
with a lepton tensor, Eq.(\ref{Lepton_tensor}), which is symmetric in its indices, the 
contribution to
the cross-section distribution, effectively, comes only from the imaginary part. Keeping the symmetry
of the lepton tensor in mind, we write
\begin{equation}
	H_{\nu\mu}^{RS}=- 4 \Im(G_1G_2^\star)Q_\mu\epsilon(p_2p_1l_1)_\nu .\label{HSO}  
\end{equation}

The same reasoning leads to the formula
\begin{equation}
	H_{\nu\mu}^{SR}=-4\Im(G_1G_2^\star)Q_\mu\epsilon(p_2p_1l_2)_\nu .\label{HOS}  
\end{equation}
Expressions (\ref{HSO}) and (\ref{HOS}) are related by the substitutions
$(p_1,l_1)\rightleftharpoons (p_2,l_2)$.

The tensors discussed so far have a  simple form. The double-spin part is rather more
complicated so we write it as a sum of several terms,

\begin{equation}
	H_{\nu\mu}^{SS}=|G_1|^2A^{11}_{\nu\mu} +G_1G_2^\star A^{21}_{\nu\mu} +
	  G_1^\star G_2A^{12}_{\nu\mu} +|G_2|^2A^{22}_{\nu\mu},\label{HadronTA}
\end{equation}
with
\begin{eqnarray}
	A^{11}_{\nu\mu}&=& -2\bigg( p_1\cdot l_1 p_2\cdot l_2+M^2l_1\cdot l_2\bigg) 
	\bigg[P_\nu  P_\mu-P^2 g_{\nu\mu} - Q_\nu Q_\mu\bigg]
	   \nonumber \\
	   && -4M^2\bigg[\half P^2 ( l_{1\mu}l_{2\nu}+l_{1\nu}l_{2\mu})
	     -P\cdot l_2 ( l_{1\mu}p_{2\nu}+l_{1\nu}p_{2\mu} ) \nonumber \\
	     &&\qquad\qquad -P\cdot l_1 ( l_{2\mu}p_{1\nu}+l_{2\nu}p_{1\mu})
	      +P\cdot l_1 P\cdot l_2g_{\nu\mu}\bigg],  \label{Aelva}
\end{eqnarray}
\begin{eqnarray}
	A^{21}_{\nu\mu}&=&- 2 M^2\big[ Q_\nu Q_\mu  l_1\cdot l_2  -
	 Q_\mu  l_{1\nu}P\cdot l_2+  Q_\mu  l_{2\nu}P\cdot l_1          \big] \nonumber \\
	   && +2 Q_\mu \bigg[p_{1\nu} p_2\cdot l_2 p_2\cdot l_1 -
	     p_{2\nu} p_1\cdot l_1 p_1\cdot l_2 -
	     \frac{1}{2} l_{1\nu}p_2\cdot l_2 P^2 +\frac{1}{2}l_{2\nu}p_1\cdot l_1P^2\bigg]  \nonumber \\
	     &=&	A^{12}_{\mu\nu} ,\label{Asymm}
\end{eqnarray}
and
\begin{equation}
	A^{22}_{\nu\mu}= - Q_\nu Q_\mu \bigg[ \frac{ Q^2 }{2M^2} (  p_1\cdot l_1 p_2\cdot l_2 -M^2l_1\cdot l_2 )
	+Q\cdot l_1 Q\cdot l_2 \bigg].
\end{equation}

There are alternative ways of formulating the above expressions. Eq.(\ref{Asymm}) could have 
been written as  
\begin{eqnarray}
	A^{21}_{\nu\mu}&=&- 2 M^2\big[ Q_\nu Q_\mu  l_1\cdot l_2  -
	 Q_\mu  l_{1\nu}Q\cdot l_2-  Q_\mu  l_{2\nu}Q\cdot l_1          \big] \nonumber \\
	   && +2 Q_\mu \bigg[p_{1\nu} p_2\cdot l_2 p_2\cdot l_1 -
	     p_{2\nu} p_1\cdot l_1 p_1\cdot l_2 +
	     \frac{1}{2} l_{1\nu}p_2\cdot l_2 Q^2 -\frac{1}{2}l_{2\nu}p_1\cdot l_1Q^2\bigg]  .
\end{eqnarray}

Now, from the symmetry expressed in Eq.(\ref{Asymm}) it follows that the imaginary part of
the hadronic tensor of Eq.(\ref{HadronTA})
 vanishes. Furthermore, if we take into consideration that the
hadronic tensor is contracted with a symmetric lepton tensor we may write
Eq.(\ref{HadronTA}) as
\begin{equation}
	H_{\nu\mu}^{SS}=|G_1|^2A^{11}_{\nu\mu} +2\Re(G_1G_2^\star) A^{21}_{\nu\mu} +
	 |G_2|^2A^{22}_{\nu\mu}.\label{HadronTA_new}
\end{equation}

%
%
%%%%%%%%%%%%%%%%%%%%%%%%%%%%%%%%%%
  \newpage
  \section{Cross-section distribution}

The next step in the calculation is the contraction of hadronic and leptonic tensors.
 The reduced cross-section distribution is defined as
\begin{equation}
	\overline{|{\cal{M}}_{red}|^2}=L^{\mu\nu}H_{\mu\nu},
\end{equation}
 and we decompose  the right hans side as
\begin{equation}
	\overline{|{\cal{M}}_{red}|^2} =\bar{R}_\Lambda R_\Lambda M^{RR}+\bar{R}_\Lambda S_\Lambda M^{RS}
	   +\bar{S}_\Lambda R_\Lambda M^{SR} +\bar{S}_\Lambda S_\Lambda M ^{SS}.\label{M-decomp}
\end{equation}
From the structure of the lepton tensor of Eq.(\ref{Lepton_tensor}), we conclude that 
each of the $M$ functions has two parts,
\begin{equation}
	M=\frac{1}{y_1y_2}\bigg[-4P^2 A -(2sP^2+y_1^2+y_2^2)B         \bigg].
\end{equation}
The $A$ factor is obtained by contracting the hadron tensor with the symmetric tensor
$	k_{1\mu}k_{1\nu} + k_{2\mu}k_{2\nu}$, 
and the $B$ factor by contracting the hadron tensor with the tensor $ g_{\mu\nu}.$ Remember that terms in the lepton tensor
containing $P_\mu$ or $P_\nu$ do not give any contribution due to the gauge invariance of the 
hadronic tensor.

The leading term of Eq.(\ref{M-decomp}) is  $M^{RR}$ and, it is independent of variables that relate to
spin dependence in 
the hyperon decay distributions. We have
\begin{eqnarray}
	A^{RR} &=& 2|G_1|^2\bigg[ (k_1\cdot P)^2+(k_2\cdot P)^2-(k_1\cdot Q)^2-(k_2\cdot Q)^2\bigg] \nonumber\\
	&& +4\Re(G_1G_2^\star) \bigg[(k_1\cdot Q)^2+(k_2\cdot Q)^2\bigg]  
	 - |G_2|^2 \frac{Q^2}{2M^2}\bigg[(k_1\cdot Q)^2+(k_2\cdot Q)^2\bigg] , \label{eqARR}
\end{eqnarray}
and 
\begin{eqnarray}
	B^{RR} &=& -4|G_1|^2( P^2+2M^2)
	 +4\Re(G_1G_2^\star) Q^2  -|G_2|^2 \frac{(Q^2)^2}{2M^2}.\label{eqBRR}
\end{eqnarray}
Thus, the distribution function $M ^{RR}$ does not depend on anyone of the decay momenta 
$l$ or $q$ of the Lambda hyperons.

Next in order are terms linear in the spin variables,
\begin{eqnarray}
	A^{RS} &=& -4\Im(G_1G_2^\star)  
	   \bigg[   k_1\cdot Q\, \mbox{det}(p_2p_1l_1k_1) +  k_2\cdot Q\, \mbox{det}(p_2p_1l_1k_2)\bigg]  ,\label{eqARS}\\
	  A^{SR} &=& -4 \Im(G_1G_2^\star)  
	   \bigg[   k_1\cdot Q\, \mbox{det}(p_2p_1l_2k_1) +  k_2\cdot Q\, \mbox{det}(p_2p_1l_2k_2)\bigg] ,\label{eqASR}
\end{eqnarray}
with $\mbox{det}(abcd)=\epsilon_{\alpha\beta\gamma\delta}a^\alpha b^\beta c^\gamma d^\delta$ and 
\begin{eqnarray}
	B^{RS}& =& 0, \label{eqBRS} \\	B^{SR} &= &0\label{eqBSR}	.
\end{eqnarray}

The expressions for the spin-spin contributions are more complicated. We have for the $A$ term
\begin{eqnarray}
	A^{SS} &=& - 2|G_1|^2 \bigg[   \bigg((k_1\cdot P)^2+(k_2\cdot P)^2-(k_1\cdot Q)^2-(k_2\cdot Q)^2\bigg)
	 \bigg(p_1\cdot l_1 p_2\cdot l_2+M^2l_1\cdot l_2\bigg) \nonumber \\
	 && \qquad \quad +2M^2\bigg(  P^2(k_1\cdot l_1 k_1\cdot l_2  +k_2\cdot l_1 k_2\cdot l_2) 
	  -2P\cdot l_2(k_1\cdot l_1 k_1\cdot p_2 +k_2\cdot l_1 k_2\cdot p_2 )\nonumber \\
	&& \qquad \qquad\qquad\quad -2P\cdot l_1(k_1\cdot l_2 k_1\cdot p_1 +k_2\cdot l_2 k_2\cdot p_1 )\bigg)
	  \bigg] \nonumber \\
	  &&-4\Re(G_1G_2^\star)\bigg[ M^2l_1\cdot l_2\bigg( (k_1\cdot Q)^2+(k_2\cdot Q)^2 \bigg)
	  \nonumber \\
	 && \qquad \quad - M^2\bigg( P\cdot l_2(k_1\cdot Q k_1\cdot l_1 + k_2\cdot Q k_2\cdot l_1 ) 
	 -P\cdot l_1(k_1\cdot Q k_1\cdot l_2 + k_2\cdot Q k_2\cdot l_2 )\bigg)  \nonumber \\
	 &&\qquad \quad -k_1\cdot Q\bigg(k_1\cdot p_1 p_2\cdot l_1p_2\cdot l_2-k_1\cdot p_2 p_1\cdot l_1p_1\cdot l_2-
	   \half P^2 k_1\cdot l_1p_2\cdot l_2 +\half P^2 k_1\cdot l_2 p_1\cdot l_1  \bigg)\nonumber \\
	 &&\qquad \quad -k_2\cdot Q\bigg(k_2\cdot p_1 p_2\cdot l_1p_2\cdot l_2-k_2\cdot p_2 p_1\cdot l_1p_1\cdot l_2-
	   \half P^2 k_2\cdot l_1p_2\cdot l_2 +\half P^2 k_2\cdot l_2 p_1\cdot l_1  \bigg) \bigg]\nonumber \\
	   &&-|G_2|^2 \frac{1}{2M^2}\bigg( (k_1\cdot Q)^2+(k_2\cdot Q)^2\bigg)\bigg[ Q^2
	     \bigg( p_1\cdot l_1p_2\cdot l_2-M^2l_1\cdot l_2\bigg)
	     +2M^2Q\cdot l_1 Q\cdot l_2 \bigg]  , \label{eqASS}
\end{eqnarray}
and for the $B$ term 
\begin{eqnarray}
	B^{SS} &=&+4|G_1|^2 \bigg[  (P^2+2M^2)(p_1\cdot l_1 p_2\cdot l_2 +M^2 l_1\cdot l_2) \nonumber \\
	 && \qquad\qquad  -M^2\bigg( P^2l_1\cdot l_2 +2 P\cdot l_2 l_1\cdot p_1
	       +2 P\cdot l_1 l_2\cdot p_2\bigg)\bigg]     \nonumber \\
	  &&  -4\Re(G_1G_2^\star)\bigg[ Q^2 M^2 l_1\cdot l_2 -
	     M^2\bigg( Q\cdot l_1P\cdot l_2-Q\cdot l_2P\cdot l_1 \bigg)     \nonumber \\
	 &&   \qquad \quad -\bigg( p_1\cdot Q p_2\cdot l_1 p_2\cdot l_2 -   p_2\cdot Q   p_1\cdot l_1  p_1\cdot l_2
	        -\half P^2 Q\cdot l_1 p_2\cdot l_2 +\half P^2 Q\cdot l_2 p_1\cdot l_1             \bigg) \bigg] \nonumber \\
	&&  -|G_2|^2 \frac{Q^2}{2M^2} \bigg[Q^2
	     \bigg( p_1\cdot l_1p_2\cdot l_2-M^2l_1\cdot l_2 \bigg)
	     +2M^2Q\cdot l_1 Q\cdot l_2 \bigg].  \label{eqBSS}
\end{eqnarray}

The functions $A^{SS}$ and $B^{SS}$ describe the joint-decay distributions of the Lambda and anti-Lambda 
hyperons. The distributions are entangled, i.e.\ they cannot be written as a product of Lambda and 
anti-Lambda distribution functions. As can be seen, even factors of the type $l_1\cdot l_2$ appear in the
joint-decay distribution. In addition, our distribution functions are explicitly covariant, as they are
expressed in terms of the four-momentum vectors of the participating particles. It is not necessary  to transform to other
coordinate systems, as in Refs.\ \cite{BaBar} and \cite{Czyz}. Another important point is that our
calculation correctly counts  the number of intermediate hyperon states.

%
%
%
%%%%%%%%%%%%%%%%%%%%%%%%%%%%%%%%%%
  \newpage
  \section{Discussion}

The distributions presented so far refer to distributions in the momenta of the hyperon
decay products. It might be of interest to integrate, say over the 
proton and pion momenta of the Lambda hyperon. To do this we need
to perform the integral over $l_{1\mu}$. This is done with recourse to  the formula
\begin{equation}
  l_\mu  \,	\textrm{dLips}(p;l,q)   = \frac{p\cdot l}{M^2}\,p_\mu \,\textrm{dLips}(p;l,q),
\end{equation}
where $p^2=M^2$, and $p\cdot l$  constant. Thus, the 
 effect of the integration is equivalent to making the substitution
\begin{equation}
	l_\mu\rightarrow \left[\frac{M^2+m^2-\mu^2}{2M^2}\right]p_\mu.\label{substitute}
\end{equation}
The phase-space volume is $l_\Lambda/4\pi M$, with $l_\Lambda$ the decay momentum in 
the Lambda rest system.

The functions $A^{RR}$ and  $B^{RR}$ of Eqs.(\ref{eqARR}) and (\ref{eqBRR}) do not depend on the variable $l_1$, but 
$A^{RS}$ of Eq.(\ref{eqARS}) does. Evidently, making the substitution (\ref{substitute}) one obtains
\begin{equation}
	A^{RS}(l_1\rightarrow p_1)=0.
\end{equation}
The other terms that depend on $l_1$ are $A^{SS}$ of Eq.(\ref{eqASS}) and  $B^{SS}$ of Eq.(\ref{eqASS}). Similarly, also here a
substitution  gives 
\begin{eqnarray}
	A^{SS}(l_1\rightarrow p_1)&=&0, \\
	B^{SS}(l_1\rightarrow p_1)&=&0.
\end{eqnarray}
This result is important since it shows that the lifetime of the Lambda
hyperon does not depend on the parameter $S_\Lambda$, and hence is  independent of the production mechanism.

Upon integration over the decay distributions of both hyperons we get
\begin{equation}
	\rd \sigma(e^+ e^- \rightarrow \gamma \Lambda \bar{\Lambda})= 
	\left[ \frac{(4\pi \alpha)^3}{2s(P^2)^2} \frac{\Gamma^2_\Lambda}{\Gamma^2} \right]  M^{RR}\,
	   \textrm{dLips}(k_1+k_2;q,p_1,p_2),	   
\end{equation}
with widths $\Gamma=\Gamma(\Lambda\rightarrow all)$ and $\Gamma_\Lambda=\Gamma(\Lambda\rightarrow p\pi^-)$,
and cross-section distribution
\begin{equation}
		M^{RR}=\frac{1}{y_1y_2}\bigg[-4P^2 A^{RR} -(2sP^2+y_1^2+y_2^2)B^{RR}\bigg] , 
\end{equation}
and with $A^{RR}$ and $B^{RR}$ as in Eqs.(\ref{eqARR}) and (\ref{eqBRR}).

%
%
%
%%%%%%%%%%%%%%%%%%%%%%%%%%%%%%%%%%
  \newpage
  \section{Folding method}
  We shall now demonstrate that the folding method used in Refs.\cite{Novo} and \cite{Czyz} 
  for calculating cross-section distributions indeed gives the same result as the present,
  conventional  method.
  To this end we need some properties of the Lambda-four-spin vector $	s(p,n)$ of  Appendix A,
\begin{equation}
		s(p,n)=\bigg( \frac{\mathbf{n}\cdot \mathbf{p}}{M},\ \frac{E\mathbf{n}\cdot \hat{\mathbf{p}}}{M}\hat{\mathbf{p}}
		+ \mathbf{n} - \hat{\mathbf{p}} ( \mathbf{n} \cdot \hat{\mathbf{p}})\bigg),
\end{equation}
where the three-vector   $\mathbf{n}$ identifies the quatization direction of the spin in the Lambda rest system.
For each $\mathbf{n}$ there are two spin states, represented by $s(p,n)$ and  $-s(p,n)$.
 
 We assume all quantization directions $\mathbf{n}$  equally likely and define averages such that
\begin{equation}
	\left\langle 1 \right\rangle =1, \qquad
	\left\langle n^k \right\rangle =0,  \qquad
	\left\langle n^k n^l \right\rangle =\delta^{kl}.	
\end{equation}
It is not difficult to show that these relations imply $\left\langle s^\mu (p,n)\right\rangle=0$, and
\begin{equation}
	\left\langle s^\mu (p,n) s^\nu (p,n)\right\rangle= \frac{1}{M^2}p^\mu p^\nu -g^{\mu\nu},
	\label{spin-tensor-average}
\end{equation} 
conditions which are explicitly covariant.

In the present investigation   cross-section distributions are obtained  by squaring the sum of the two matrix elements
corresponding to  the  diagrams of Fig.1, i.e.\ by calculating
\begin{equation}
	\left| {\cal{M}}(e^+ e^- \rightarrow \gamma \Lambda(\rightarrow p\pi^-) \bar{\Lambda}(\rightarrow \bar{p}\pi^+))\right|^2.
\end{equation}
The matrix element of a diagram is a product of a hyperon-production step and subsequent hyperon-decay steps,
with sums over the intemediate hyperon-spin states. This is embodied in the hadron tensor of Eq.(\ref{hadron-tensor-trace}).

In the folding method of Refs.\cite{Novo} and \cite{Czyz}  one first calculates  cross-section and decay distributions for
given hyperon spins, and then averages their product over spin-quantization directions according to Eq.(\ref{spin-tensor-average}). 
Thus,  the prescription is to form
\begin{equation}
\left\langle 	\left| {\cal{M}}(e^+ e^- \rightarrow \gamma \Lambda_n \bar{\Lambda}_{n'})\right|^2
 \left| {\cal{M}}(\Lambda_n \rightarrow  p\pi^-)\right|^2
\left| {\cal{M}}(\bar{\Lambda}_{n'} \rightarrow \bar{p}\pi^+)\right|^2  \right\rangle_{nn'}. \label{fold_I}
\end{equation}
In addition, we should multiply   by a factor of four, since for each quantization direction 
there are two spin possibilities, spin up and spin down. Details of the calculation are
given in Appendix B.

The squared matrix element for Lambda decay when summed over final proton spin states 
is, as  in Appendix A,
\begin{equation}
	\left| {\cal{M}}(\Lambda_n \rightarrow  p\pi^-)\right|^2=R_\Lambda +M S_\Lambda l_1\cdot s_1 ,
\end{equation}
and the squared matrix element for hyperon  production 
$\left| {\cal{M}}(e^+ e^- \rightarrow \gamma \Lambda_n \bar{\Lambda}_n)\right|^2$
contains the projector
\begin{equation}
	u(p_1,s_1)\bar{u}(p_1,s_1)=(\slashed p_1+M)\half (1+\gamma_5 \slashed s_1 ),
\end{equation}
with $s_1=s(p_1,n).$ Multiplying the product of these two expressions by the factor of two,
 for the two spin possibilities,
 and taking the average according to Eq.(	\ref{spin-tensor-average}), it follows that 
\begin{equation}
	2 \left\langle (\slashed p_1+M)\half (1+\gamma_5 \slashed s_1 )\bigg[ R_\Lambda +M S_\Lambda l_1\cdot s_1\bigg] \right\rangle
	= (\slashed p_1+M)  \bigg[ R_\Lambda  -  S_\Lambda \gamma_5(p_1\cdot l_1 +M\slashed l _1  ) \bigg].
\end{equation}
This result is immediately recognized as the Lambda-hyperon factor $X_\Lambda$ of Eq.(\ref{LamXfactor}) in the trace 
form of the hadronic tensor, Eq.(\ref{hadron-tensor-trace}).

For the anti-Lambda hyperon the projector is
\begin{equation}
	v(p_2,s_2)\bar{v}(p_2,s_2)=(\slashed p_2 - M)\half (1+\gamma_5 \slashed s_2 ),
\end{equation}
and combined with the anti-Lambda-decay distribution
\begin{equation}
	\bar{R}_\Lambda +M \bar{S}_\Lambda l_2\cdot s_2,
\end{equation}
it leads to the average 
\begin{equation}
	2 \left\langle (\slashed p_2-M)\half (1+\gamma_5 \slashed s_2 )
	   \bigg[ \bar{R}_\Lambda +M \bar{S}_\Lambda l_2\cdot s_2\bigg] \right\rangle
	= (\slashed p_2- M)  \bigg[\bar{ R}_\Lambda  +  \bar{S}_\Lambda \gamma_5(p_2\cdot l_2  - M\slashed l _2  ) \bigg],
\end{equation}
a result identical to the $Y_\Lambda$ factor of Eq.(\ref{anLamYfactor}) which describes the anti-Lambda-hyperon 
factor of the hadronic tensor, Eq.(\ref{hadron-tensor-trace}).

We conclude that the folding method as used in Refs.\cite{Novo} and \cite{Czyz}  leads to  the same
result as a conventional evaluation of  Feynman diagrams, provided the number of spin states is
correctly counted. 

In the conventional calculation there is correlation, or entanglement, between the hyperon decay products
already in the matrix element. In the cross-section distribution, e.g., this is manifested in
the term $l_1\cdot l_2$. Moreover, the matrix element involves a sum over intermediate
hyperon polarizations, but once we have chosen the spin-quantization direction, there are
only two contributions, spin up and spin down, as is clear from the decomposition
\begin{equation}
	\slashed p+M=\sum_{s=\pm} u(p,s)\bar{u}(p,s)=(\slashed p+M)\half (1+\gamma_5 \slashed s )
	+(\slashed p+M)\half (1-\gamma_5 \slashed s ),
\end{equation}
where $s=s(p,n)$ with $\vec{n}$ arbitrary but fixed. Thus, only one quantization direction is 
considered and the result is independent of the one chosen.

In the folding calculation of Eq.(\ref{fold_I}) one starts with a product of distribution functions
for fixed quantization directions, $n$ and $n'$. As a consequence, the cross-section-distribution 
function 
factorizes into a product of distribution functions. This implies vanishing correlation between the
decay products of the two hyperons. However, taking the average of a product distribution 
over the quantization directions $n$ and $n'$
does not necessarily yield a product distribution. Instead correlations between the various
factors are created, in such a way as to reproduce the correct result.
%
%
%%%%%%%%%%%%%%%%%%%%%%%%%%%%%%%%%%
  \newpage
  \begin{acknowledgments}
I thank Bengt Karlsson, Stefan Leupold, and Karin Sch\"onning for discussions.

\end{acknowledgments}
%%%%%%%%%%%%%%%%%%%%%%%%%%%%%%%%%%%%%%%%%%%%%%%%%%%%%%%%%%%%%%%
%
%
\newpage
\appendix
\section{}

The matrix element for $\Lambda\rightarrow p \pi^-$ decay is commonly written as
\begin{equation}
	{\cal M}(\Lambda\rightarrow p \pi^-)=\bar{u}_p(l)[ A+B\gamma_5]u_\Lambda(p). \label{decay-L}
\end{equation}
For a Lambda hyperon, of polarization $+\mathbf{n}$ in its rest system, 
the square of this matrix element, after summation over final-state-proton 
polarizations, becomes
\begin{equation}
\sum|{\cal M}|^2=\mbox{Sp}[(A^\star-B^\star\gamma_5)(\slashed l+m)(A+B\gamma_5)
   (\slashed p+M)\half(1+\gamma_5 \slashed s)] .
 \label{decay-L2}
\end{equation}
The spin four-vector $s=s(p,n)$ satisfies $s\cdot s=-1$ and $s\cdot p=0$, 
where $p$ is the hyperon four-momentum. The mass of the hyperon is $M$ and
that of the  proton  $m$. The spin vector for polarisation  $-\mathbf{n}$ 
is $-s(p,n)$.

In the rest system of the Lambda $s(p,\mathbf{n})=(0,\mathbf{n})$ and 
$p=(M,\mathbf{0})$. In a coordinate system where the Lambda has
three-momentum $\mathbf{p}$, the spin vector is
\begin{equation}
	s(p,n)=\frac{n_{\parallel}}{M}(|\mathbf{p}|, E \hat{\mathbf{p}})+(0,\mathbf{n}_\bot),
	\label{spin-vector}
\end{equation}
with $n_{\parallel}=\mathbf{n}\cdot\hat{\mathbf{p}}$ and
\begin{equation}
	\mathbf{n}_\bot=\mathbf{n} -\hat{\mathbf{p}}(\mathbf{n}\cdot \hat{\mathbf{p}}).  
\end{equation}
In the first part of expression $(\ref{spin-vector})$ we notice the helicity
vector $h(p)=(|\mathbf{p}|, E \hat{\mathbf{p}})/M$.

Evaluation of the trace gives the  distribution function
\begin{equation}
\sum|{\cal M}|^2=R_\Lambda +M S_\Lambda l\cdot s ,
 \label{decay-L3}
\end{equation}
where
\begin{eqnarray}
R_\Lambda &=& |A|^2((M+m)^2-\mu^2) + |B|^2((M-m)^2-\mu^2) , \\
S_\Lambda &=& 4\Re (A^\star B) ,
 \label{RandS}
\end{eqnarray}
and $\mu$ the pion mass. In the rest system of the Lambda the decay distribution  is
\begin{eqnarray}
	\sum|{\cal M}|^2&=&R_\Lambda (1+ \alpha_\Lambda\hat{\mathbf{l}}\cdot \mathbf{n}) ,\\
	\alpha_\Lambda&=&\frac{-l_\Lambda MS_\Lambda}{R_\Lambda},\\
	l_\Lambda&=& \frac{1}{2M}\bigg[((M+m)^2-\mu^2)((M-m)^2-\mu^2)\bigg]^{1/2},\label{decayl}
\end{eqnarray}
with $l_\Lambda$ the decay momentum in the Lambda rest system.

For unpolarized decay we average over
the two spin vectors $s(p,n)$ and $-s(p,n)$, and get  
\begin{equation}
|{\cal M}|^2_{unpol}=R_\Lambda .
\end{equation}
The decay width is
\begin{equation}
	\Gamma_\Lambda(\Lambda \rightarrow p \pi^- )= \frac{l_\Lambda}{8\pi M^2}R_\Lambda ,
\end{equation}
with $l_\Lambda$ as in Eq.(\ref{decayl}).

The matrix element for the  charge conjugate decay, $\bar{\Lambda}\rightarrow \bar{p} \pi^+$, is 
\begin{equation}
	{\cal M}(\bar{\Lambda}\rightarrow \bar{p} \pi^+)=\bar{v}_\Lambda(p)[ A'+B'\gamma_5]v_p(l), \label{decay-Lbar}
\end{equation}
and the corresponding  decay-distribution function
\begin{equation}
\sum|{\cal M}|^2=\bar{R}_\Lambda +M \bar{S}_\Lambda l\cdot s ,
 \label{decay-L2bar}
\end{equation}
where
\begin{eqnarray}
\bar{R}_\Lambda &=& |A'|^2((M+m)^2-\mu^2) + |B'|^2((M-m)^2-\mu^2),\\
\bar{S}_\Lambda &=& 4\Re (A'^\star B') .
 \label{ARandS}
\end{eqnarray}
 For unpolarized decay of anti-Lambda, $ |{\cal M}|^2_{unpol}=\bar{R}_\Lambda$
 .
CP invariance implies  $A=A'$ and $B=-B'$, so that $\alpha_{\bar{\Lambda}}=-\alpha_{\Lambda}$.

%%%%%%%%%%%%%%%%%%%%%%%%%%%%%%%%%%%%%%%%%%%%%%%%%%%%%%%%%%%%%%%
%
%
\newpage
\section{}
In order to facilitate comparison with  Czy\.z et al., \cite{Czyz}, we have  calculated the hadronic tensor
$K_{\nu\mu}(s_1,s_2)$
for the reaction $e^+ e^- \rightarrow \gamma \Lambda \bar{\Lambda}$. 
Here, the spin vector for the final state Lambda is denoted $s_1$ and for the
anti-Lambda $s_2$, both referring to spin up. The hadronic tensor is defined by
\begin{equation}
	K_{\nu\mu}(s_1,s_2)= \mbox{Sp}[ \bar{O}_\nu (\slashed p_1+M)\half(1+\gamma_5 \slashed s _1) 
	   O_\mu  (\slashed p_2-M)\half(1+\gamma_5 \slashed s _2) ], \label{LLbar-tensor}
\end{equation}
with the matrix $ O_\mu $ is  in Eq.(\ref{Omudef}). This hadronic tensor is gauge invariant,
and is decomposed as
\begin{equation}
	K_{\nu\mu}(s_1,s_2)=K_{\nu\mu}^{00}(0,0) + K_{\nu\mu}^{05}(s_1,0) + K_{\nu\mu}^{50}(0,s_2) + K_{\nu\mu}^{55}(s_1,s_2).
	\label{LLtensdiv}
\end{equation}
The functional arguments indicate the spin vectors involved. 

The first term on the right hand side has been calculated before, and 
\begin{equation}
	K_{\nu\mu}^{00}(0,0) =\frac{1}{4} H_{\nu\mu}^{RR},
\end{equation}
with $H_{\nu\mu}^{RR}$ defined in Eq.(\ref{hOO}). Since the leptonic tensor is symmetric
in its indices we need only retain the symmetric part of the hadronic tensor. It follows that
\begin{eqnarray}
	K_{\nu\mu}^{05}(s_1,0) &=& \frac{-1}{2M}\Im(G_1G_2^\star)\bigg[Q_\nu\epsilon(p_1,p_2,s_1)_\mu
	    +Q_\mu\epsilon(p_1,p_2,s_1)_\nu \bigg] ,  \\
	K_{\nu\mu}^{50}(0,s_2) &=& \frac{-1}{2M}\Im(G_1G_2^\star)\bigg[Q_\nu\epsilon(p_1,p_2,s_2)_\mu
	    +Q_\mu\epsilon(p_1,p_2,s_2)_\nu \bigg]    ,
\end{eqnarray}
with the epsilon function  of Eq.(\ref{epsI}).

The  contribution depending on both spin vectors is  
\begin{equation}
K_{\nu\mu}^{55}(s_1,s_2)=|G_1|^2B^{1}_{\nu\mu} +|G_2|^2B^{2}_{\nu\mu}
    +\Re(G_1G_2^\star) B^{3}_{\nu\mu} ,
\end{equation}
with
\begin{eqnarray}
	B^{1}_{\nu\mu}&=&- \bigg[ p_{1\nu} p_{2\mu}  +p_{1\mu}  p_{2\nu} -\half g_{\nu\mu}P^2 \bigg] s_1\cdot s_2
	    -\half P^2 ( s_{1\nu}   s_{2\mu} + s_{1\mu}   s_{2\nu} )   \nonumber \\
	     &&+p_1\cdot s_2 ( s_{1\nu}p_{2\mu}+s_{1\mu}p_{2\nu})
	     +p_2\cdot s_1 ( s_{2\nu}p_{1\mu}+s_{2\mu}p_{1\nu})
	      -g_{\nu\mu}p_1\cdot s_2    p_2\cdot s_1       ,\\
	B^{2}_{\nu\mu}&=& \frac{1}{4M^2} Q_\nu Q_\mu \bigg[ \half Q^2 s_1\cdot s_2+p_1\cdot s_2 p_2\cdot s_1 \bigg],\\
   	B^{3}_{\nu\mu}&=& -  Q_\nu Q_\mu  s_1\cdot s_2 +
      \half \bigg[ p_1\cdot s_2( Q_\nu  s_{1\mu} +Q_\mu  s_{1\nu}) - p_2\cdot s_1
          (Q_\nu  s_{2\mu} +  Q_\mu s_{2\nu}) \bigg].       
\end{eqnarray}
This hadronic tensor corresponds to the one of Czy\.z et al., Eq.(7) of Ref. \cite{Czyz}, but written on a 
covariant form.

Next, we fold the $\Lambda\bar{\Lambda}$  tensor  with  the 
decay distributions of the hyperons, i.e.\ we  first multiply  by 
\begin{equation}
	4\bigg[ R_\Lambda +M S_\Lambda l_1\cdot s_1\bigg]\bigg[ \bar{R}_\Lambda +M \bar{S}_\Lambda l_2\cdot s_2\bigg]
\end{equation}
and then average over spin diections according to Eq.(\ref{spin-tensor-average}). 
Remember that the factor of four is needed 
to properly include the total number of spin states. The directional average  involves 
\begin{equation}
	\left\langle s_1(l_1\cdot s_1)\right\rangle =\frac{p_1\cdot l_1}{M^2}p_1 -l_1.
\end{equation}
Thus, the hadronic tensor $	K_{\nu\mu}(s_1,s_2)$ of Eq.(\ref{LLbar-tensor}) is replaced by the folded tensor
\begin{eqnarray}
	H_{\nu\mu}&=&4 \bigg[ \bar{R}_\Lambda R_\Lambda  K_{\nu\mu}^{00}(0,0) 
	  - M  \bar{R}_\Lambda S_\Lambda   K_{\nu\mu}^{05}(l_1,0) -  M  \bar{S}_\Lambda R_\Lambda  K_{\nu\mu}^{50}(0,l_2) \nonumber \\
	 && +M^2 \bar{S}_\Lambda S_\Lambda \bigg\{ 
	  K_{\nu\mu}^{55}(l_1 ,l_2 ) 
	  -  \frac{p_1\cdot l_1}{M^2} K_{\nu\mu}^{55}(p_1,l_2)-  \frac{p_2\cdot l_2}{M^2} K_{\nu\mu}^{55}(l_1, p_2) \nonumber \\
	 && \qquad \qquad \qquad + \frac{p_1\cdot l_1}{M^2} \frac{p_2\cdot l_2}{M^2} K_{\nu\mu}^{55}(p_1,p_2)
	    \bigg\} \bigg],
\end{eqnarray}
which agrees, term by term, with the  tensor of Eq.(\ref{hadron-exp}), but is a factor of four bigger
than that of   Czy\.z et al.\ \cite{Czyz}.
\newpage
\section{}
Suppose we integrate over the anti-Lambda decay distribution. Then the reduced cross-section distribution
 of Eq.(\ref{M-decomp}) is reduced to
\begin{equation}
	\overline{|{\cal{M}}_{red}|^2} =\bar{R}_\Lambda\bigg[ R_\Lambda M^{RR}+ S_\Lambda M^{RS}\bigg].
\end{equation}
The function $M^{RS}$ has only a contribution from $A^{RS}$ of Eq.(\ref{eqARS}), which has the form of 
a four-spin vector $S_1$ contracted with the proton four-momentum vector $l_1$. We define  $S_1$ from
\begin{equation}
	  k_1\cdot Q\epsilon(p_2p_1l_1k_1) +  k_2\cdot Q\epsilon(p_2p_1l_1k_2)=S_1\cdot l_1,
\end{equation}
From $p_1\cdot S_1=0$ it follows that $S_1$ is a typical spin vector, except for the fact it is not properly normalized.
This is easily arranged for, by defining 
\begin{equation}
	s_{1\mu}(p_1)=\frac{1}{D}S_{1\mu}=
	\frac{1}{D}\bigg[k_1\cdot Q\epsilon(p_1p_2k_1)_\mu +  k_2\cdot Q\epsilon(p_1p_2k_2)_\mu\bigg],
\end{equation}
with $S_1\cdot S_1=-D^2$, and
\begin{eqnarray}
	D^2 &=& (k_1\cdot Q)^2\left|
%\begin{vmatrix}
\begin{array}{ccc}
	p_1\cdot p_1 & p_1\cdot p_2 & p_1\cdot k_1 \\
	p_2\cdot p_1 & p_2\cdot p_2 & p_2\cdot k_1 \\
	k_1\cdot p_1 & k_1\cdot p_2 & k_1\cdot k_1 
%\end{vmatrix}
\end{array}\right|
+(k_2\cdot Q)^2\left|
\begin{array}{ccc}
	p_1\cdot p_1 & p_1\cdot p_2 & p_1\cdot k_2 \\
	p_2\cdot p_1 & p_2\cdot p_2 & p_2\cdot k_2 \\
	k_2\cdot p_1 & k_2\cdot p_2 & k_2\cdot k_2 
\end{array}\right|  \nonumber \\
&& +2k_1\cdot Q k_2\cdot Q \left|
\begin{array}{ccc}
	p_1\cdot p_1 & p_1\cdot p_2 & p_1\cdot k_1 \\
	p_2\cdot p_1 & p_2\cdot p_2 & p_2\cdot k_1 \\
	k_2\cdot p_1 & k_2\cdot p_2 & k_2\cdot k_1 
\end{array}\right|
\end{eqnarray}

We can now summarize the Lambda decay distribution as
\begin{equation}
	\overline{|{\cal{M}}_{red}|^2} =\bar{R}_\Lambda R_\Lambda M^{RR}
	\bigg[1 - P_\Lambda \alpha_\Lambda l_1\cdot s_1/l_\Lambda \bigg],
\end{equation}
with the Lambda polarization 
\begin{equation}
	P_\Lambda=\frac{-16 \Im(G_1G_2^\star)P^2D/M}{4P^2 A^{RR}+(2sP^2+y_1^2+y_2^2)B^{RR}},   
\end{equation}
and $l_\Lambda$ the decay momentum in the Lambda rest system. If we go to the Lorentz system 
where the Lambda hyperon is at rest, then the normalized spin vector reads
\begin{equation}
	\vec{s}_1(p_1)=\frac{M}{D}\bigg[ k_1\cdot Q \vec{p}_2\times\vec{k}_1 +k_2\cdot Q \vec{p}_2\times\vec{k}_2\bigg].
\end{equation}

%%%%%%%%%%%%%%%%%%%%%%%%%%%%%%%%%%

\end{document}